\documentclass[prl,reprint,superscriptaddress,aps,floatfix]{revtex4-2}
\usepackage{amsmath}
\usepackage{amsfonts}
\usepackage{graphicx}
\usepackage{physics}
\usepackage[usenames]{color}

\begin{document}
\date{\today}

\title{Counterdiabatic Route to Entanglement Steering and Dynamical Freezing in the Floquet Lipkin-Meshkov-Glick Model}
\author{Nakshatra Gangopadhay}
\author{Sayan Choudhury}
\email{sayanchoudhury@hri.res.in}
\affiliation{Harish-Chandra Research Institute, a CI of Homi Bhabha National Institute, Chhatnag Road, Jhunsi, Allahabad 211019}
	
\begin{abstract}
Controlling the dynamics of quantum many-body systems is crucial for developing quantum technologies. This work demonstrates that counter-diabatic (CD) driving provides a powerful tool for steering collective spin systems along entangled trajectories for a long time. In particular, CD driving leads to approximate stroboscopic freezing and eternal entanglement oscillations for a large class of initial states in the periodically driven Lipkin-Meshkov-Glick model. Intriguingly, CD driving generates spin squeezing and its associated metrologically useful multipartite entanglement at the mid-point of every drive cycle, when the system is initially prepared in a fully x-polarized state. The CD driving induced non-ergodic dynamics is accompanied by a decrease in the average eigenstate entanglement and inverse participation ratio, thereby signalling greater eigenstate localization. Our work opens a new route to evade Floquet heating and control entanglement generation in collective spin systems.

\end{abstract}
	
\maketitle

{\it Introduction:} Periodic driving provides a powerful tool for tailoring the behavior of complex quantum systems~\cite{bukov2015universal,weitenberg2021tailoring,oka2019floquet}. On the one hand, it presents a route to engineer interesting effective Hamiltonians that describe the stroboscopic evolution of the system~\cite{rudner2020floquet,rudner2020band,eckardt2017colloquium,banerjee2024emergent}. On the other hand, these driving protocols can be employed to steer many-body systems along a desired trajectory, thereby enabling the realization of non-equilibrium phases of matter such as anomalous topological insulators~\cite{harper2020topology,moessner2021topological} and time crystals~\cite{sacha2018review,sacha2020book,khemani2019review,nayak2019review,zaletel2023colloquium}. Unfortunately, driving inevitably leads to infinite-temperature thermalization, thereby posing a major challenge to the coherent control of periodically driven (Floquet) systems at long times~\cite{d2014long,choudhury2014stability,bukov2015prethermal,mallayya2019heating,ikeda2021fermi}.

Several quantum control techniques have been devised to steer many-body systems along a desired trajectory at short times~\cite{demirplak2003adiabatic,van2016optimal,sauvage2020optimal,maskara2021discrete,ljubotina2022optimal,tomsovic2023controlling,ljubotina2024tangent,beringer2024controlling}. In particular, quantum annealing has been successfully employed to prepare quantum states by adiabatically changing the Hamiltonian parameters~\cite{das2008colloquium,albash2018adiabatic,hauke2020perspectives}. In recent years, `shortcut-to-adiabaticity' (STA) techniques have also been developed to accelerate quantum annealing~\cite{torrontegui2013shortcuts,jarzynski2013generating,deffner2014classical,guery2019shortcuts,hatomura2024shortcuts,campbell2015shortcut,guery2023driving,hatomura2017shortcuts,mc2024towards}. These advances naturally raise a tantalizing question: can STA protocols be harnessed to control the dynamics of Floquet matter over a long time?

We affirmatively answer this question by demonstrating that a widely used STA protocol - counter-diabatic (CD) driving - can provide a route to steer Floquet collective spin systems through an entangled trajectory for long times. CD driving employs an auxiliary CD Hamiltonian, that coherently removes diabatic excitations between the eigenstates of the original Hamiltonian~\cite{berry2009transitionless,del2013shortcuts,sels2017minimizing,claeys2019floquet,morawetz2024efficient,passarelli2020counterdiabatic,zhou2024approximate,zhou2020experimental,nakahara2022counterdiabatic,takahashi2024shortcuts,schindler2024counterdiabatic,vcepaite2023counterdiabatic,barone2024counterdiabatic,hegade2021shortcuts,chandarana2022digitized,kumar2021counterdiabatic,kadowaki2023greedy}. While a closed-form expression of the CD Hamiltonian can be derived~\cite{berry2009transitionless}, it is usually very cumbersome to implement exact CD driving in a many-body system. Intriguingly, some recent works have proposed systematic strategies to obtain local CD driving protocols that can be experimentally implemented, enabling fast state preparation~\cite{sels2017minimizing,claeys2019floquet,morawetz2024efficient}.

\begin{figure}
    \centering
    \includegraphics[width=0.48\textwidth]{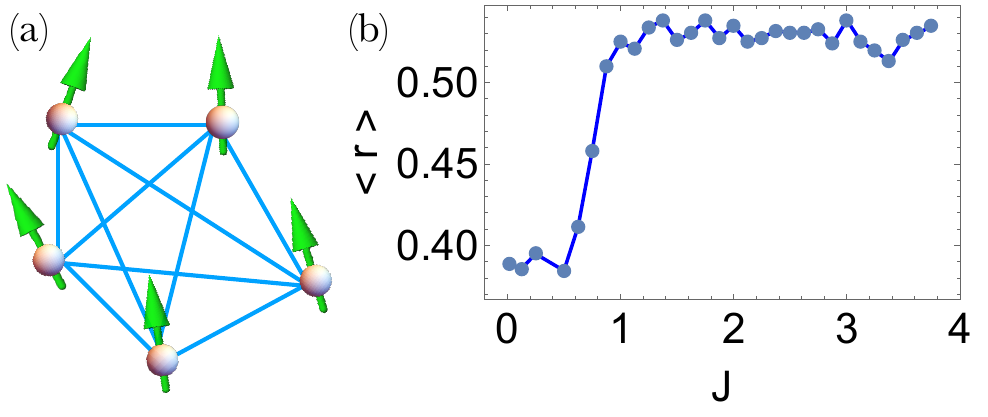}
    \caption{{\bf Model and Spectral Statistics:} (a) A schematic illustration of the Lipkin-Meskov-Glick (LMG) model that is characterized by uniform `all-to-all' interactions amongst all the spin-$1/2$ particles. (b) The averaged level spacing ratio, $\langle r \rangle$ of the driven LMG model as a function of the interaction strength $J$ for $N=2000$. The system exhibits a transition from integrable behavior with Poisson statistics $(\langle r \rangle \approx 0.39)$ to chaotic behavior with Wigner-Dyson statistics $(\langle r \rangle \approx 0.53)$, when $J \sim 1$.}
    \label{fig:illustration}
\end{figure}

In this letter, we investigate the effect of adding local CD driving terms to the Floquet Lipkin-Meshkov-Glick (LMG) model~\cite{lipkin1965validity,ribeiro2007thermodynamical,li2023improving}. The LMG model is characterized by uniform all-to-all interactions between spin-$1/2$ particles (see fig.~\ref{fig:illustration}(a)), and it naturally describes the dynamics of atoms in collective cavity quantum-electrodynamics systems~\cite{hosten2016quantum,muniz2020exploring}. Intriguingly, we identify a wide parameter regime, where the system can exhibit eternal entanglement oscillations accompanied by approximate many-body freezing at stroboscopic times. We trace the origin of this dynamical freezing to CD-driving-induced eigenstate localization. Our results demonstrate that CD driving can be a powerful tool to control entanglement in Floquet systems, leading to potential applications in quantum computing and metrology.  \\

{\it CD Driving:} We now outline the technique of CD driving that we employ in this work; a more detailed description can be found in ref~\cite{suppmat}. Let us consider a Hamiltonian $H(\lambda)$, where $\lambda$ is time-dependent. The usual method of adiabatic state preparation involves changing $\lambda$ adiabatically from $0$ to $1$, such that the system always remains in the ground state of $H(\lambda)$; this method can be implemented in any gapped system.

The key idea of CD driving is to change $\lambda(t)$ at a finite rate, while simultaneously mitigating diabatic excitations by adding a suitable counter term: 
\begin{equation}
H_{\rm CD}(t) = H(\lambda)+\dot{\lambda}\mathcal{A}_\lambda,
\end{equation}
where $\mathcal{A}_\lambda$ is the adiabatic gauge potential (AGP)~\cite{berry2009transitionless}. The exact form of the AGP is
\begin{equation}
    \mathcal{A}_{\lambda} = i \sum_n \left(\vert \partial_t n\rangle\langle n \vert - \langle n \vert \partial_t n \rangle \vert n \rangle \langle n \vert \right) = \sum_{k=1}^{\infty} \alpha_k \mathcal{O}_{\rm LCD}^{(k)},
\end{equation}
where $\{\mathcal{O}_{\rm LCD}\}$ denote a set of Krylov space operators that implement local CD driving~\cite{claeys2019floquet,takahashi2024shortcuts} (see also ref.~\cite{hastings2004locality,hastings2005quasiadiabatic}):
\begin{equation}
\mathcal{O}_{\rm LCD}^{(k)}= i [\underbrace{H,[H,\ldots[H}_\textrm{2k-1},\partial_\lambda H]]]
\end{equation}

In this work, we use an approximate form of the AGP determined by a set of coefficients $\{\alpha_1,\alpha_2, \ldots\alpha_l\}$, where $l$ is finite and it controls the locality of the Hamiltonian. In particular, we focus on the $l=1$ (dubbed `CD1'), and $l=2$ (dubbed `CD2') cases, and determine $\alpha_l$ by minimizing the action $\mathcal{S}_l = \Tr[\mathcal{G}^2_l]$, where $\mathcal{G}_l = \left(\partial_\lambda H - i[H,\mathcal{A}_\lambda^{(l)}]\right)$~\cite{sels2017minimizing,claeys2019floquet}. The utility of this technique for suppressing diabatic excitations in quantum annealing has already been extensively investigated. We now proceed to examine its efficacy in controlling Floquet heating.\\

\begin{figure}
    \centering
    \includegraphics[width=0.47\textwidth]{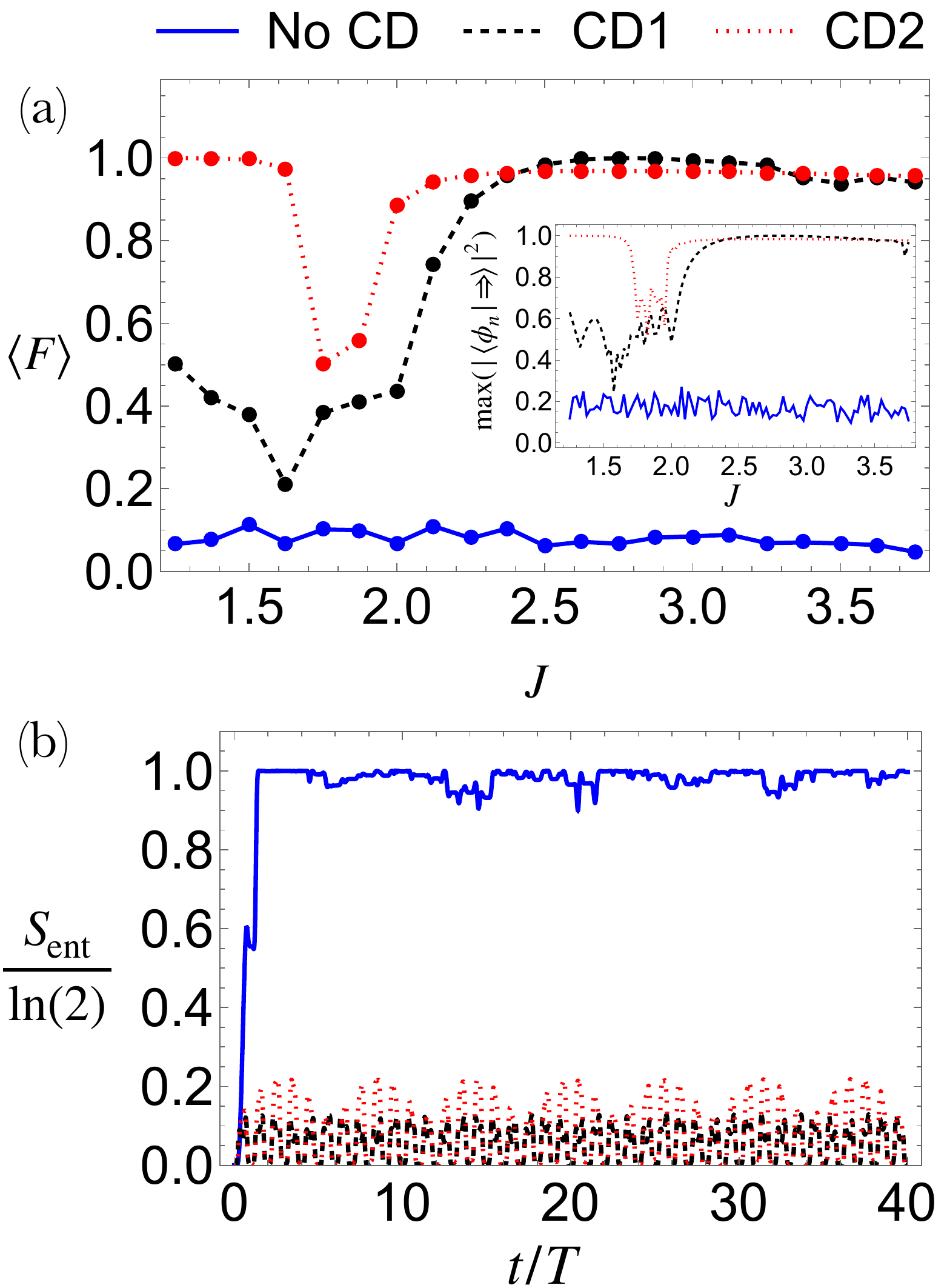}
    \caption{{\bf Dynamics of the x-polarized initial state:} (a) The stroboscopic fidelity, $F = \vert \langle \psi(nT)|\psi(0) \rangle \vert^2$ averaged over $2000$ periods as a function of $J$.  In the absence of CD driving, the system quickly thermalizes and loses all memory of initial conditions. However, the system exhibits persistence of memory when CD driving is implemented. For CD driving, $\langle F \rangle$ exhibits a non-monotonic dependence on $J$ and it shows approximate stroboscopic freezing when $J>2.5$. The inset shows the behavior of the maximum overlap of the x-polarized state $\vert\!\Rightarrow\!\rangle$ with the Floquet eigenstates. The stroboscopic freezing can be traced to the existence of a localized eigenstate with a large overlap with $\vert\!\Rightarrow\!\rangle$. (b) The time-evolution of the entanglement entropy $S_{\rm ent}$ (see text for definition) when $J=3.125$. The system thermalizes in the absence of CD driving $\left (S_{\rm ent} \sim \ln(2) \right)$. Intriguingly, both CD driving protocols lead to eternal entanglement oscillations indicating that the system is steered through an entangled trajectory. These results have been obtained for $N=100$.}
    \label{fig:dynamics}
\end{figure}

{\it Model:} We will now examine the dynamics of the Floquet LMG model described by the Hamiltonian:
\begin{equation}
    H = -\bigl(1-\lambda(t)\bigr) \sum_{i=1}^N \sigma_i^x + \lambda(t) \frac{J}{N} \sum_{i,j=1}^N \sigma_i^z \sigma_j^z,
\label{eq:model}
\end{equation}
where we set $J>0$. This system is widely used for quantum metrology since it can be employed to prepare spin-squeezed states~\cite{kitagawa1993squeezed,sorensen2001entanglement,ma2011quantum}. It is worth noting that $H$ is SU$(2)$ invariant and the total spin, $S^2={\bf S.S}$ is conserved; here $S^{\alpha} = 1/2 \sum_i \sigma_i^{\alpha}$. In this work, we examine the system's time-evolution from states initially prepared in the $S=\vert {\bf S} \vert = N/2$ sector; all fully polarized initial states belong to this category. This choice constrains the time-evolution to a $(N+1)$-dimensional subspace thereby enabling us to study large system sizes.

In order to compare our results to the extant CD driving results, we shall work with a specific driving protocol $\lambda(t)=\sin^2{\left[\left(\frac{\pi}{2}\right)\sin^2{}\left(\frac{\pi t}{2 \tau}\right)\right]}$, such that the drive period $T$ is $2 \tau$~\cite{sels2017minimizing,claeys2019floquet,passarelli2020counterdiabatic,morawetz2024efficient}. In this case, the ground state of $H$ at $t=0$ is the fully x-polarized state: $\vert\!\Rightarrow\!\rangle = \vert\!\rightarrow \rightarrow \ldots \rightarrow\!\rangle$ state. When the system is initially prepared in the $\vert\!\Rightarrow\!\rangle$ state, an adiabatic state preparation would lead to a highly entangled Dicke State: $ \vert S^z =0 \rangle $ at $t= \tau$~\cite{opatrny2016counterdiabatic}; the system would return to the unentangled $\vert\!\Rightarrow\!\rangle$ state at $t=2 \tau$. Thus, the system is steered along an entangled trajectory, when $\tau \rightarrow \infty$. We now proceed to study the dynamics of this system when $\tau$ is finite ($\tau=1$)  and analyze the effect of the CD1 and CD2 protocols. CD1 driving is implemented using $\mathcal{O}^{(1)}_{LCD} = \frac{8J}{N
}\left(S_y S_z + S_zS_y\right)$. Implementing the CD2 protocol also requires $\mathcal{O}^{(2)}_{\rm LCD}$, which has a more cumbersome form~\cite{suppmat}. These driving protocols can be experimentally realized using the procedure outlined in ref.~\cite{claeys2019floquet}.\\

{\it Stroboscopic freezing and entanglement oscillations:} Before looking at the effect of CD driving, we first analyze the spectral statistics of the Floquet Hamiltonian. To do this, we determine the Floquet operator, $U_F = \mathcal{T} \exp[-i \int dt H(t)] = \exp(- i H_F T) = \sum_n \exp(-i \epsilon^F_{n} T) \vert \phi_n \rangle \langle \phi_n \vert $, {where $\epsilon^F_{n}$ represent the quasienergies of the Floquet Hamiltonian, $H_F$~\cite{pal2010many,d2014long,ponte2015periodically,zhang2016floquet}. We then determine the level-spacing ratios, $r_n = {\rm min}(d_n, d_{n+1})/ {\rm max}(d_n, d_{n+1}) $, where $d_n = \epsilon^F_{n+1}-\epsilon^F_{n}$ is the Floquet level-spacing. It is well-known that when a system is integrable (chaotic), the spectral statistics is Poisson (Wigner-Dyson) with $\langle r \rangle \sim 0.39 \,\, (\langle r \rangle \sim 0.53)$~\cite{atas2013distribution,giraud2022probing}; a more detailed discussion of the spectral statistics is presented in \cite{suppmat}. As shown in Fig.~\ref{fig:illustration}(b), this system transitions from integrable to chaotic behavior when $J \sim 1$. In this work, we focus on the $J>1$ regime, when the system is ergodic and explore the effect of CD driving on the long-time dynamics. 

We now proceed to study the dynamics of the system when it is initially prepared in the $\vert\!\Rightarrow\!\rangle$ state. As discussed earlier, this process would steer the system along an entangled trajectory in the adiabatic limit. We compute the stroboscopic return probability $F = \vert \langle \psi(nT)\vert \psi(0) \rangle \vert^2$ and find that the system completely loses the memory of its initial conditions in the chaotic regime when CD driving is absent. However, the situation changes dramatically in the presence of CD driving. In particular, $F$ is always enhanced by CD driving; however, this enhancement has a non-monotonic dependence on $J$. At smaller values of $J$, CD2 leads to a significantly higher value of $F$ compared to CD1 and $F$ decreases with increasing $J$. However, when $J \gtrsim 2.5$, both CD1 and CD2 lead to approximate stroboscopic freezing. We trace the origin of this freezing to a large overlap of $\vert\!\Rightarrow\!\rangle$ with a localized Floquet eigenstate. Our results are shown in fig.~\ref{fig:dynamics}(a). We note that this kind of dynamical many-body freezing usually requires strong driving~\cite{das2010exotic,bhattacharyya2012transverse,hegde2014freezing,haldar2018onset,haldar2021dynamical,haldar2022statistical,haldar2017dynamical} or multi-tone protocols~\cite{banerjee2024exact}.

\begin{figure}
    \centering
    \includegraphics[width=0.47\textwidth]{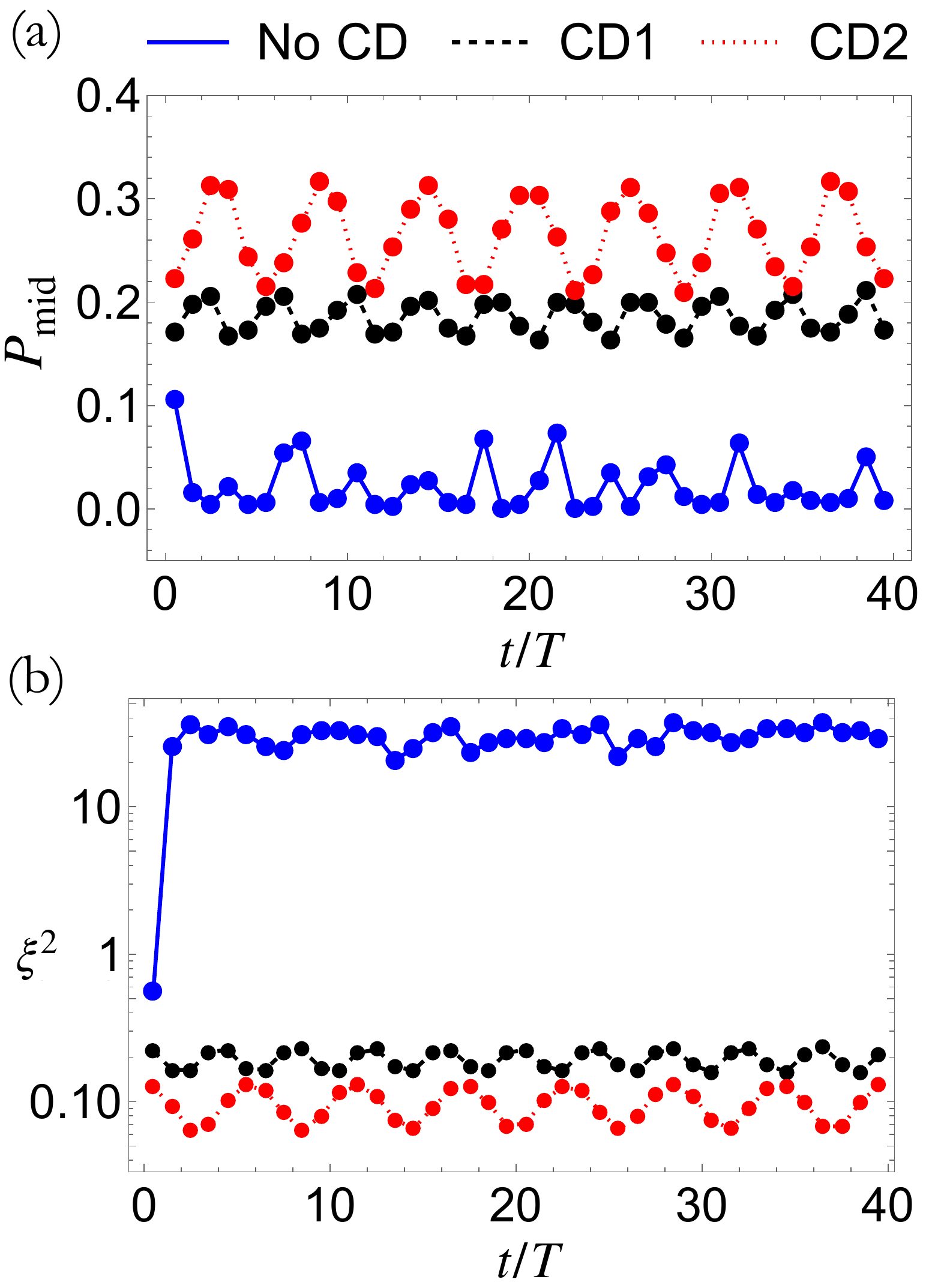}
    \caption{{\bf Entanglement Steering:} (a) The time-evolution of the overlap of the state at the middle of each drive cycle with the target Dicke state, $P_{\rm mid} = \vert \langle \psi\bigl((n+1/2)T\bigr) \vert  S^z =0 \rangle \vert^2$. CD driving leads to significant enhancement of $P_{\rm mid}$. (b) The time-evolution of the spin-squeezing parameter, $\xi^2 = \frac{\Delta S_z^2}{N/4}$. CD driving leads to the generation of spin squeezing and multipartite entanglement characterized by $\xi^2 < 1$. These results have been obtained for $N=100$ and $J=3.125$.}
    \label{fig:squeeze}
\end{figure}

Having established the persistence of memory in the presence of CD driving, we examine the entanglement entropy generated during the evolution in the regime where CD driving leads to approximate stroboscopic freezing. Due to the collective nature of the model, we bipartition one spin from the remaining system and then obtain its reduced density matrix: $\rho = \left(\frac{1}{2} \mathbb{I} + \frac{1}{N}\sum_{\alpha} \langle S^{\alpha} \rangle \sigma^{\alpha}\right)$. We then compute the entanglement entropy $S_{\rm ent} = \rho \ln(\rho)$; this entropy has been measured experimentally~\cite{neill2016ergodic}. We find that in the absence of CD driving, $S_{\rm ent}$ quickly grows to its maximum possible value of $\ln(2)$. Intriguingly, in the presence of CD driving, the system exhibits eternal entanglement oscillations, where $S_{\rm ent}$ grows during the micromotion and then returns to $0$ at the end of each period. Our conclusions remain unchanged even if other bipartitions are considered~\cite{suppmat}. 

\begin{figure}[t]
    \centering
    \includegraphics[width=0.45\textwidth]{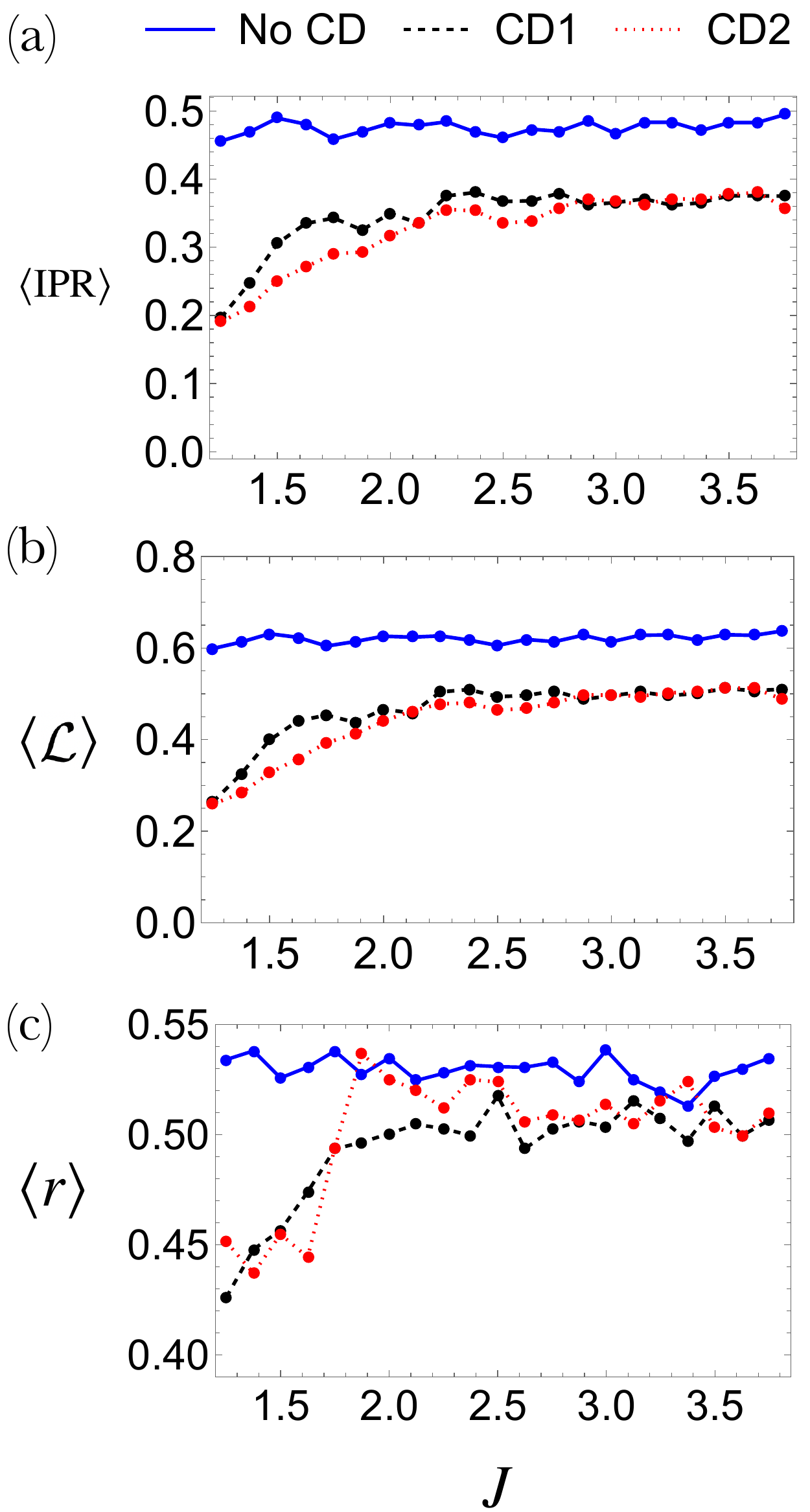}
    \caption{{\bf CD driving induced Eigenstate Localization:} (a) The eigenstate averaged Inverse participation ratio (IPR) is lowered in the presence of CD driving, thereby indicating greater localization. (b) The eigenstate averaged Wehrl entropy localization measure, $\mathcal{L}$ of the eigenstates is decreased by CD driving. Both the IPR and $\mathcal{L}$ of the eigenstates indicate that CD2 induces greater localization than CD1 when $J$ is small; at larger values of $J$, their effect is similar. (c) The spectral statistics of this system is captured by the averaged level spacing ratio, $\langle r \rangle$. In the presence of CD driving, $\langle r \rangle$ takes a value lower than the Wigner-Dyson prediction of $\langle r \rangle \sim 0.53$. For these calculations, $N$ has been set to $100$ for (a) and (b) and $2000$ for (c).}
    \label{fig:eig}
\end{figure}

We further investigate the dynamics of the system by examining the overlap of the state at the mid-point of every drive cycle, $\vert \psi\bigl((n+1/2) T \bigr) \rangle$ with the $\vert S^z=0 \rangle$ Dicke state,  $P_{\rm mid} = \vert \langle \psi((n+1/2)T) \vert S^z=0 \rangle|^2$, where $n \in \mathbb{Z}$. As shown in Fig.~\ref{fig:squeeze}(a), CD driving considerably enhances this overlap. Interestingly, $P_{\rm mid} (t)$ exhibits a non-monotonic behavior with time - it increases with the number of drive cycles at short times and then oscillates. Thus, if our primary goal was the fast preparation of the $\vert S^z=0 \rangle$ Dicke state with local CD driving, then a better fidelity can be obtained by employing the Floquet protocol for a few cycles, instead of the usual practice of stopping at the mid-point of the first drive cycle~\cite{sels2017minimizing,claeys2019floquet}. Furthermore, CD driving leads to the generation of significant spin squeezing, as captured by the generalized spin-squeezing parameter, $\xi^2 = \frac{\Delta S_z^2}{N/4}$~\cite{opatrny2016counterdiabatic,santra2024squeezing}. CD protocols lead to $\xi^2<1$ - a clear signature of spin-squeezing and its associated metrologically useful multipartite entanglement~\cite{sorensen2001many,sorensen2001entanglement,toth2009spin,lucke2014detecting,opatrny2016counterdiabatic,suppmat} (see Fig.~\ref{fig:squeeze}(b)). We conclude that CD driving can steer collective spin systems along an entangled trajectory.\\  

\begin{figure}[b]
    \centering
    \includegraphics[width=0.48\textwidth]{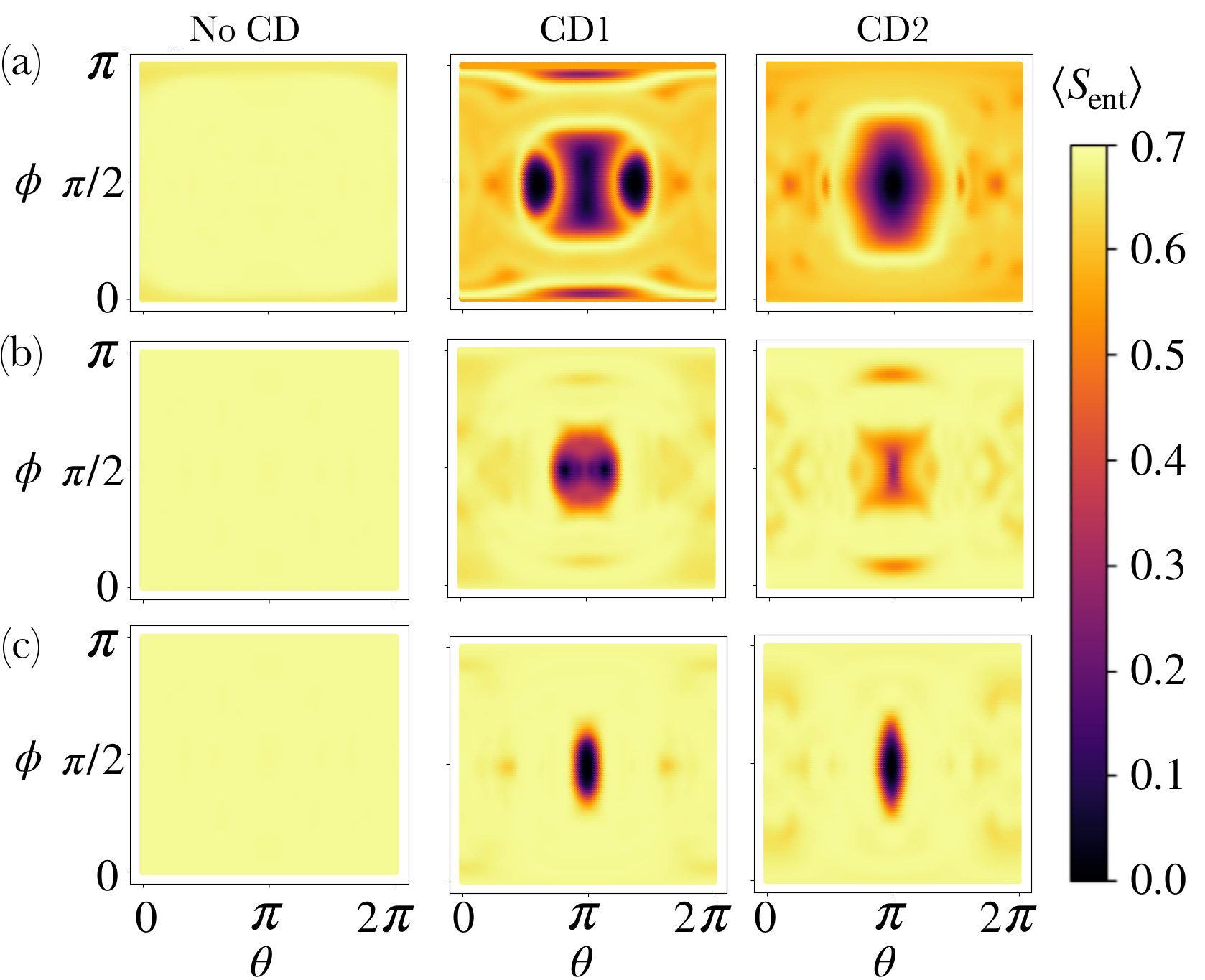}
    \caption{{\bf Local Chaos:} The long-time stroboscopic entanglement entropy, $S_{\rm ent}$ for initial spin-coherent states, where $S_{\rm ent} (nT)$ has been averaged over 1000 oscillations between $n=9500$ to $n=10500$ for (a) $J=1.25$, (b) $J=1.875$, and (c) $J=3.125$ and $N=100$. In the absence of CD driving, the system is ergodic and $\langle S_{\rm ent} \rangle \sim \ln(2)$. However, in the presence of CD driving there is a large class of initial conditions for which the system evades thermalization $(\langle S_{\rm ent} \rangle \sim 0)$.}
    \label{fig:chaos}
\end{figure}

{\it Eigenstate localization and local chaos:} Since CD driving can have a significant impact on the quench dynamics, it is natural to investigate its influence on the entire eigenspectrum. We do this by first examining the behavior of two eigenstate averaged phase-space localization measures - the Inverse participation ratio (IPR) and the Wehrl entropy localization measure ($\mathcal{L}$)~\cite{wang2023statistics,yan2024addendum}; more details about these measures is provided in ref.~\cite{suppmat}. Both of these quantities are based on the Husimi function $Q(\theta, \phi)$. The Husimi function of a Floquet eigenstate, $\vert \mu_n \rangle$ is defined as $Q_n (\theta, \phi)=\vert \langle \theta, \phi \vert \mu_n \rangle \vert^2$, where $\vert \theta, \phi \rangle$ are the generalized $\rm{SU(2)}$ spin-coherent states: 
\begin{equation}
\vert \theta, \phi \rangle = \exp\left[i \theta(S^x \sin(\phi) - S^y \cos(\phi))\right] \vert S^z = N/2 \rangle.
\end{equation}
The IPR for the Floquet eigenstate $\vert \mu_n \rangle$, $I_n$ is then defined as:
\begin{equation}
I_n = \frac{(N+1)^2}{4 \pi}\left[\int d\phi d\theta \sin(\theta) Q_n^2(\theta,\phi)\right]^{-1}.
\end{equation}
$I_n$ is very small ($\sim 0$) for extremely localized eigenstates and it takes a value of $1$ for a fully delocalized eigenstate. As shown in fig.~\ref{fig:eig}(a), in the ergodic regime, the eigenstate-averaged IPR reaches the same value as other chaotic collective systems such as the kicked top ($\langle {\rm IPR} \rangle \sim 0.5$)~\cite{wang2023statistics}. Interestingly, CD driving considerably reduces $\langle {\rm IPR} \rangle$, indicating that these protocols induce greater localization.

A related measure that can be employed to examine this non-ergodicity further is the Wehrl entropy~\cite{wehrl1979relation}:
\begin{equation}
    S_w =  - \frac{N+1}{4 \pi}\int d\phi d\theta \sin(\theta) Q_n(\theta,\phi)\ln[Q_n(\theta,\phi)].
\end{equation}
$S_w$ can characterize the entanglement complexity of many-body states~\cite{sugita2003moments} and it is minimum for coherent states~\cite{lieb1978proof}. The Wehrl entropy localization measure, $\mathcal{L} = \bigl(\exp(S_w)/(N+1)\bigr)$ can efficiently capture the phase space localization~\cite{suppmat}. The eigenstate-averaged value of $\mathcal{L}$ takes the value of $0.655$ for large $N$ in a completely ergodic system~\cite{yan2024addendum}. As shown in fig.~\ref{fig:eig}(b), the system reaches this value in the absence of CD driving in the ergodic regime. However, analogous to the IPR, the value of $\langle \mathcal{L} \rangle$ is lowered significantly in the presence of CD driving. These results together conclusively demonstrate that CD driving can be employed to realize non-ergodicity in periodically driven quantum systems.

Finally, we characterize phase-space localization by examining the long-time behavior of the stroboscopic entanglement entropy $S_{\rm ent}$ for dynamics initiated from coherent states. For collective spin systems, this measure characterizes local (global) chaos in the non-ergodic (ergodic) regime~\cite{piga2019quantum}; our results are shown in fig.~\ref{fig:chaos}. In the absence of CD driving, $S_{\rm ent} \sim \ln(2)$ for all initial states, since the system is ergodic. However, CD driving leads to a large class of initial states for which $S_{\rm ent}$ remains at very small values ($\sim 0$) at stroboscopic times, and the system exhibits localization. Furthermore, we find that at smaller values of $J$, both the eigenstate localization measures ($\langle {\rm IPR} \rangle$ and $\langle \mathcal{L} \rangle$) and the class of initial states that exhibit localized dynamics are different for the CD1 and CD2 protocols. This difference almost disappears when $J \gtrsim 2.9$. These findings demonstrate that the long-time dynamics after a quench provide important insights into CD driving-induced non-ergodicity.\\

{\it Summary and outlook:} The controllable generation and manipulation of entangled states is essential for the development of quantum technologies. Unfortunately, entanglement creation is almost inevitably accompanied by thermalization in periodically driven many-body systems. This poses a major challenge to harnessing these systems for quantum information processing applications. We have demonstrated that CD driving presents a powerful route to mitigate this problem by profoundly affecting the nature of the Floquet eigenstates. This effect is particularly striking, when the system is ergodic in the absence of CD driving. Intriguingly, CD driving induces localization for a large fraction of the Floquet eigenstates in this regime. Consequently, these systems can exhibit eternal entanglement oscillations and stroboscopic freezing for a large class of initial states. Furthermore, CD driving leads to the generation of spin squeezing and its associated metrologically useful multipartite entanglement. We conclude that CD driving can be an effective technique for steering Floquet systems along an entangled trajectory without thermalizing.

This work serves as the starting point for exploring several interesting avenues for future research. A natural next step would be to investigate the dynamics of periodically driven systems when optimal control techniques are employed in conjunction with CD driving. Another fruitful direction would be to explore routes to employ CD-driven Floquet systems for quantum metrology and quantum simulation. Finally, it would be interesting to explore the effect of CD driving on open Floquet systems.

\section{Acknowledgements} 
SC thanks DST, India for support through SERB project SRG/2023/002730 and W. Vincent Liu for discussions.

\bibliographystyle{apsrev4-2}
\bibliography{ref}

\end{document}

\title{Supplemental Material: Counterdiabatic Route to Entanglement Steering and Dynamical Freezing in the Floquet Lipkin-Meshkov-Glick Model}
\author{Nakshatra Gangopadhay}
\author{Sayan Choudhury}
\email{sayanchoudhury@hri.res.in}
\affiliation{Harish-Chandra Research Institute, a CI of Homi Bhabha National Institute, Chhatnag Road, Jhunsi, Allahabad 211019}

\maketitle
The supplemental material outlines the derivation of the Counterdiabatic Driving Hamiltonian (Sec.~\ref{sec:CDHam}), the explicit form of the CD terms (Sec.~\ref{Sec:CDTerms}), additional numerical results for the entanglement dynamics and spin squeezing at short times (Sec.~\ref{sec:Dynamics}), an overview of eigenstate localization measures (Sec.~\ref{sec:Localization}), spin squeezing and its connections to entanglement (Sec.~\ref{sec:Squeeze}), and a brief discussion about level statistics (Sec.~\ref{sec:LevelStat}).

\section{Derivation of the Counterdiabatic Driving Hamiltonian}
\label{sec:CDHam}

In this section, we shall discuss and derive the counter-diabatic Hamiltonian for a clearer understanding of the formalism employed in the main text. Let us start by revisiting the adiabatic evolution of a quantum system. Consider a time-dependent Hamiltonian, $H(t)$ such that 
\begin{equation}
    H(t)\ket{\psi(t)}= E(t)\ket{\psi(t)},
\end{equation}
where $\ket{\psi(t)}$ is an instantaneous eigenstate of our Hamiltonian. It doesn't in general, serve as a solution of the time-dependent Schr\"{o}dinger equation. However, we can use it as a suitable basis that we can expand the solution of the time-dependent Schr\"{o}dinger equation, $\ket{\Psi(t)}$:
\begin{equation}
    H(t)\ket{\Psi(t)}= i\partial_t\ket{\Psi(t)}
\end{equation}
Let us consider a family of instantaneous eigenstates $\ket{n(t)}$, and write the general solution as follows,
\begin{equation}
    \ket{\Psi(t)}= \sum_{n} c_n(t)\ket{n(t)}.
\end{equation}
This immediately implies:
\begin{equation}
     \dot{c_k}(t) = \left(\frac{E_k}{i} - \langle k(t)\vert\dot{k}(t)\rangle\right)c_k -  \sum_{n\ne k} \frac{\dot{H}_{kn}(t)}{E_n-E_k} c_n(t),
\end{equation}
where, $\dot{H}_{kn}(t) = \langle k(t)\vert \dot{H}(t)\vert n(t)\rangle$.\\

The second term in the above equation contains all the information about transitions between different instantaneous eigenstates. However, if the system is driven extremely slowly, we can ignore the second term as all transitions from the initial instantaneous eigenstate are highly suppressed. This is the adiabatic approximation. The solution now, is given by:

\begin{equation}
     c_k(t) = c_k(0) e^{i\theta_k(t)} e^{i\gamma_k(t)};\,\, \theta_k(t) = \int_{0}^{t} E_k(t')dt',\,\,\,\, \gamma_k(t) = i\int_{0}^{t} \langle k(t)\vert \dot{k}(t)\rangle dt',
\end{equation}
where $\theta_k(t)$ is a dynamical phase, and $\gamma_k(t)$ is the geometric phase. Hence, under the adiabatic approximation
\begin{equation}
     \ket{\Psi_k(t)} = e^{i\theta_k(t)} e^{i\gamma_k(t)} \ket{k(t)} \approx \ket{\Psi(t)}
\end{equation}
 where, $\ket{\Psi_k(0)} = \ket{k(0)}$. \\

The adiabatic approximation is valid only when the system is driven extremely slowly; rapid driving inevitably leads to diabatic excitations. Intriguingly, it is possible to drive a system at any arbitrary rate and suppress the diabatic transitions at the same time. This can be achieved by following the procedure laid out in ref.~\cite{berry2009transitionless}. This procedure relies on adding auxiliary terms, $H_1$ to the original $H$, such that the states driven under the adiabatic approximation $\ket{\Psi_k(t)}$ are the exact evolving states under the new Hamiltonian, $H_{\rm CD}=H+H_1$. To find this new Hamiltonian, we note that any time-dependent Unitary operator $\hat{U}(t)$ is a solution to the Schrodinger equation, and thus, the new Hamiltonian can be written as,
\begin{equation}
    H_{\rm CD}(t)= i\left(\partial_t\hat{U}(t)\right)\hat{U}^{\dagger}(t)
\end{equation}
We can now choose our $\hat{U}(t)$ as follows,
\begin{equation}
    \hat{U}(t)=\sum_{n} e^{i\theta_n(t)}e^{i\gamma_n(t)}\ket{n(t)}\bra{n(0)},
\end{equation}
thereby implying
\begin{equation}
    H_{\rm CD}(t) = \sum_{n} E_n\ket{n}\bra{n}+i\sum_n\left(\ket{\dot{n}}\bra{n}-\langle n\vert \dot{n}\rangle\ket{n}\bra{n}\right).
\end{equation}
We conclude that $H_{\rm CD}=H+H_1$, where $H_1=i \sum_n\left(\ket{\dot{n}}\bra{n}-\langle n\vert \dot{n}\rangle\ket{n}\bra{n}\right)$. This counter-diabatic Hamiltonian can be written in a more convenient form as follows, 
\begin{equation}
    H_1=i\sum_{n\ne m}\sum \frac{\ket{m}\bra{m}\partial_t H(t) \ket{n}\bra{n}}{E_n-E_m}
\end{equation} 
Thus, using the procedure outlined above, we have found the Hamiltonian $H_{\rm CD}(t)$ that can generate transitionless driving. Unfortunately for a generic many-body system, this Hamiltonian is highly non-local and therefore difficult to engineer. However, it is possible to approximate such terms, which enables the suppression of diabatic excitations without losing locality. These protocols are dubbed local counterdiabatic driving (LCD). We now outline the method proposed in ref.~\cite{claeys2019floquet} to systematically derive these LCD protocols.\\

We start by noting that the auxiliary Hamiltonian, $H_1$ can be conveniently parametrized as $H_1 = \dot{\lambda}\mathcal{A}_{\lambda}$, where the adiabatic gauge potential (AGP), $\mathcal{A}_{\lambda}$ satisfies:
\begin{equation}
    \left[i\partial_{\lambda} H - [\mathcal{A}_{\lambda},H],H\right]=0
\end{equation}
It is easy to see that solving this equation is equivalent to minimizing the Hilbert-Schmidt norm of the operator:
\begin{equation}
    G_{\lambda}(\mathcal{A}_{\lambda})= \partial_{\lambda}H-i[H,\mathcal{A}_{\lambda}]
\end{equation}
with respect to $\mathcal{A}_{\lambda}$~\cite{sels2017minimizing}. This is equivalent to solving the Euler-Lagrange equations of the action 
\begin{equation}
    \mathcal{S}(\mathcal{A}_{\lambda})={\rm Tr}[G^2_{\lambda}(\mathcal{A}_{\lambda})]
\end{equation}. 
The solution to this equation leads to the following integral form of the AGP~\cite{hastings2004locality,hastings2005quasiadiabatic,takahashi2024shortcuts}
\begin{equation}
    \mathcal{A}_{\lambda} = -\frac{1}{2} \lim_\textrm{$\eta \rightarrow \infty$} \int_{-\infty}^{\infty} {\rm sgn}(s)e^{-\eta\vert s\vert} \times e^{iH(\lambda)s}\partial_{\lambda}H(\lambda)e^{-iH(\lambda)s}ds
\end{equation}
This integral expression is proportional to the operator $\partial_{\lambda}H(\lambda)$. Defining $\mathcal{L}_{\lambda}(.)=[H(\lambda),.]$, we can now perform the integral over $s$ and write, 
\begin{equation}
    \mathcal{A}_{\lambda} = -\frac{1}{2} \lim_\textrm{$\eta \rightarrow \infty$} \left(\frac{1}{\eta - i\mathcal{L}_{\lambda}} - \frac{1}{\eta + i\mathcal{L}_{\lambda}}\right) \partial_{\lambda}H(\lambda) 
\end{equation}

This formal expression motivates us to use an expansion of the form,
\begin{equation}
    \mathcal{A}_{\lambda} = i\sum_k \alpha^{nc}_k (\lambda) \mathcal{L}_{\lambda}^{2k-1}\partial_{\lambda}H(\lambda) = \sum_{k=1}^{\infty} \alpha_k \mathcal{O}_{\rm LCD}^{(k)}.
\end{equation}
As we have discussed in the main text, we determine an approximate form of the AGP by treating  $\{\alpha_1,\alpha_2, \ldots\alpha_l\}$, as variational parameters. We note that this form of the AGP remains well-defined even for chaotic many-body systems and it can be realized experimentally due to its similarity with the Magnus expansion~\cite{claeys2019floquet}. 

\section{Form of the CD Hamiltonian}
\label{Sec:CDTerms}

In this section, we derived the exact form for the CD terms that we have used in the main text. We start by recalling that CD driving is implemented by a Hamiltonian

\begin{equation}
\ H_{CD} = H + \dot{\lambda} \mathcal{A}_{\lambda}.
\end{equation}
Here $H$ is the time-dependent Hamiltonian in the absence of CD driving and $\mathcal{A}_{\lambda}$ is is the adiabatic gauge potential~\cite{berry2009transitionless}:
\begin{equation}
    \mathcal{A}_{\lambda} = i \sum_n \left(\vert \partial_t n\rangle\langle n \vert - \langle n \vert \partial_t n \rangle \vert n \rangle \langle n \vert \right) = \sum_{k=1}^{\infty} \alpha_k \mathcal{O}_{\rm LCD}^{(k)},
\end{equation}
where $\{\mathcal{O}_{\rm LCD}\}$ denote a set of Krylov space operators that implement local CD driving~\cite{claeys2019floquet,takahashi2024shortcuts}:
\begin{equation}
\mathcal{O}_{\rm LCD}^{(k)}= i [\underbrace{H,[H,\ldots[H}_\textrm{2k-1},\partial_\lambda H]]]
\end{equation}
For the CD driving protocols discussed in this work, we need $\mathcal{O}_{\rm LCD}^{(1)}$ and $\mathcal{O}_{\rm LCD}^{(2)}$. For the driven LMG model:
\begin{equation}
\ H= -(1-\lambda)\sum_{i=1}^N \sigma
_i^x + \lambda\frac{J}{N} \sum_{i,j=1}^N \sigma
_i^z ,
\end{equation}
these terms can be explicitly obtained:

\begin{equation}
\ \mathcal{O}^{(1)}_{\rm LCD} = i[\mathcal{H},\partial_{\lambda}\mathcal{H}] =\frac{8J}{N
}\left(S_y S_z + S_zS_y\right)
\end{equation}
and 

\begin{equation}
\begin{split}
\mathcal{O}^{(2)}_{\rm LCD} =\ & -(\frac{128J}{N}(1-\lambda)^2 \{S_z,S_y\} 
+ \frac{64\lambda J^2}{N^2} \left[\{\{S_y,S_z\},S_x\} + 2S_zS_xS_y + 2S_yS_xS_z \right] \\
& + \frac{128J^2}{N^2} (1-\lambda)^2 \left[\{\{S_z,S_x\},S_y\}\right] 
 + \frac{128\lambda^2J^3}{N^3} \left[\{S_y,S_z^3\} + 3S_z^2 S_yS_z + 3S_zS_yS_z^2\right]),
\end{split}
\end{equation}
where $\{\hat{A},\hat{B}\} = \hat{A}\hat{B} + \hat{B},\hat{A}$ represents the anti-commutator of $\hat{A}$ and $\hat{B}$.\\

Thus, we can implement the CD  driving protocols are implemented using the Hamiltonians:
\begin{eqnarray}
H_{\rm CD1} = H + \dot{\lambda} \alpha_1 \mathcal{O}^{(1)}_{\rm LCD}, \\
H_{\rm CD2} = H + \dot{\lambda} (\alpha_1 \mathcal{O}^{(1)}_{\rm LCD} + \alpha_2 \mathcal{O}^{(2)}_{\rm LCD}).
\end{eqnarray}
The procedure to obtain $\alpha_k$ is described in the main text (also see ref.~\cite{claeys2019floquet}). Finally, we note that these Hamiltonians can be implemented using the protocol outlined in ref.~\cite{claeys2019floquet}.

\begin{figure}
    \centering
    \includegraphics[width=0.8\textwidth]{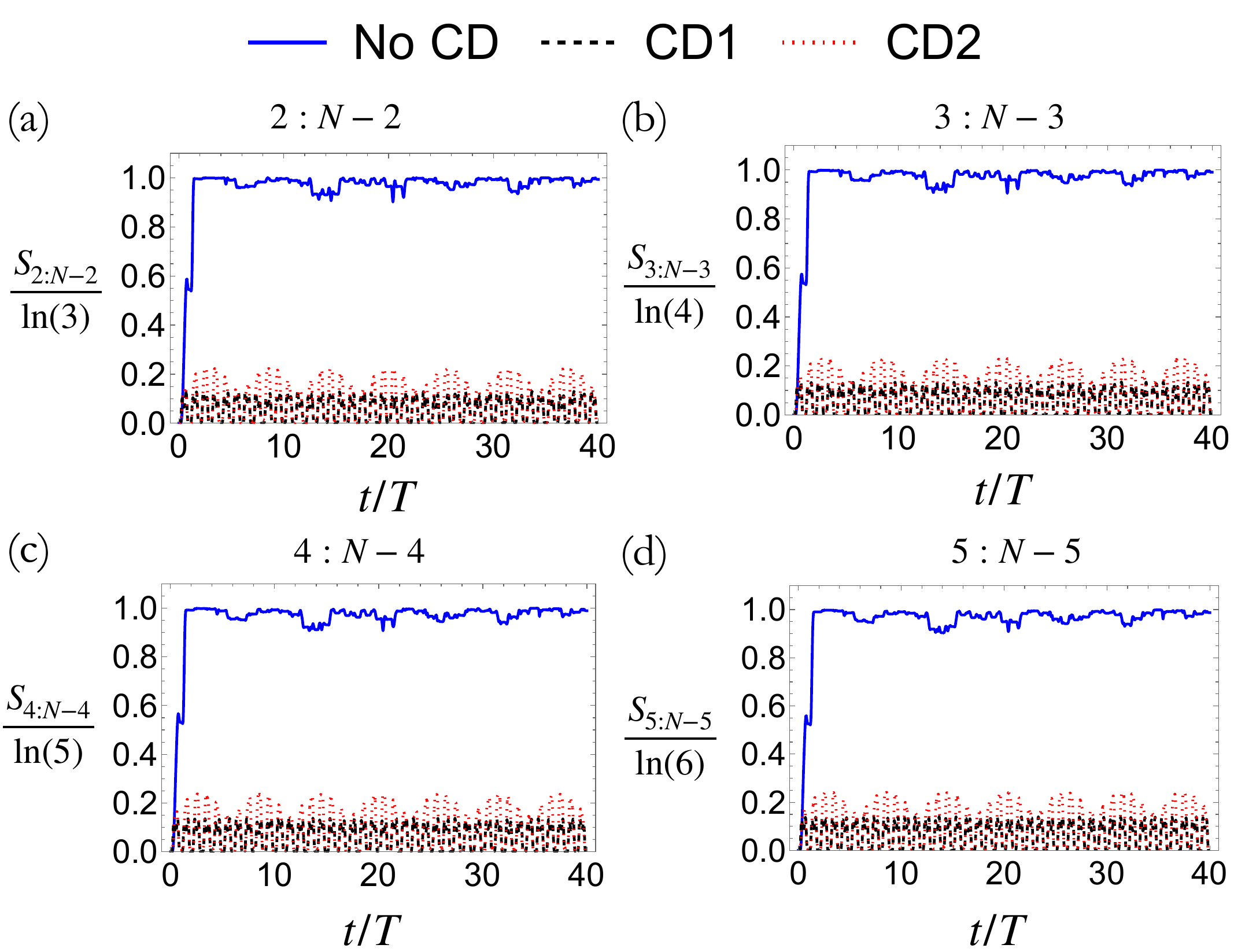}
    \caption{{\bf Entanglement entropy for different bipartitions:} The $m:N-m$ entanglement entropy, $S_{m:N-m}$ obtained by bipartitioning the system into $m$ and $N-m$ spins at $J=3.125$ for the x-polarized initial state $\vert \Rightarrow \rangle$. In the absence of CD driving, the $S_{m:N-m}$ saturates to a value of $\ln(m+1)$. However, in the presence of CD driving, the entanglement entropy exhibits eternal oscillations. Furthermore, $S_{m:N-m}/\ln(m+1)$ behaves in the same manner for all these bipartitions. $N$ has been set to 100 for these calculations.}
    \label{fig:suppfig1}
\end{figure}

\section{Short Time Dynamics}
\label{sec:Dynamics}
In this section, we discuss two aspects of the short-time dynamics of the system: (a) the dynamics of the entanglement entropy for various bipartitions of the system and (b) the efficacy of CD driving for the preparation of the Dicke state.
\subsection{Dynamics of Entanglement}
\label{subsec:DynE}
In the main text, we have discussed the time-evolution of the entanglement entropy, $S_{\rm ent}$ obtained by bipartitioning the $N-$spin LMG chain into 1 and $N-1$ spins. In that case, we found that in the absence of CD driving, $S_{\rm ent}$ quickly saturated to the maximal value of $\ln(2)$, while CD driving leads to non-ergodic dynamics with eternal entanglement oscillations. A natural question then is to explore the time-evolution of the entanglement entropy for other bipartitions of the system. In Fig.~\ref{fig:suppfig1}, we present results for the $m:N-m$ entanglement entropy, $S_{m:N-m}$, obtained by partitioning the system into $m$ and $N-m$ spins, for the x-polarized initial state, $\vert \Rightarrow \rangle$ when $J=3.125$ and $N=100$. These parameters are identical to the one used in Fig.~2(b) of the main text. We find that in this case, in the absence of CD driving, $S_{m:N-m}$ saturates to a value of $\ln(m+1)$, while the CD protocols lead to eternal entanglement oscillations. Furthermore, we note that $S_{m:N-m}/\ln(m+1)$, behaves almost identically for all protocols for the values of $m$ that we have considered ($2,3,4$ and $5$). Our results provide further evidence for the non-ergodic dynamics induced by CD driving.

\begin{figure}
    \centering
    \includegraphics[width=0.85\textwidth]{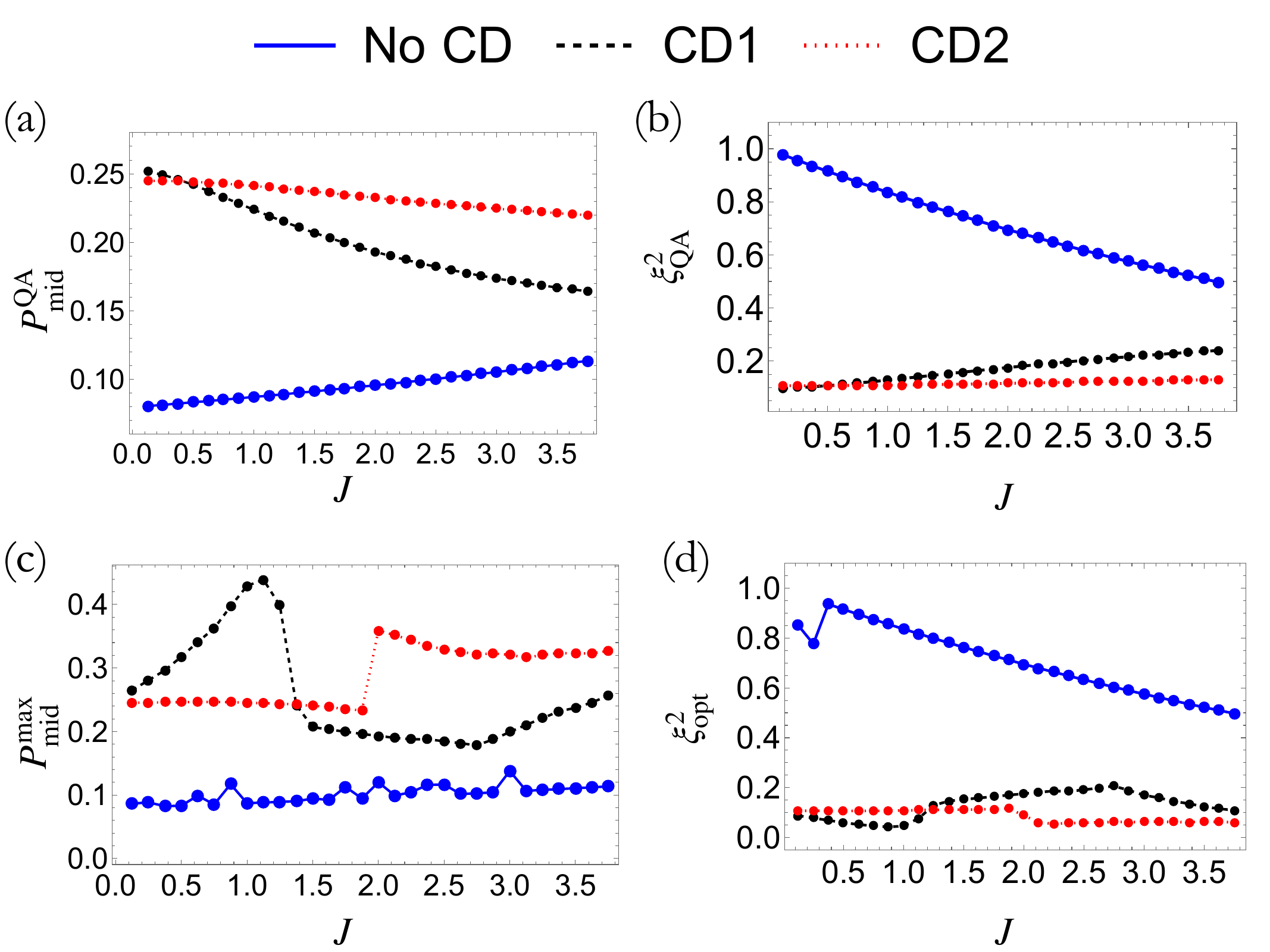}
    \caption{{\bf Dicke State Preparation with CD driving:} (a)-(b) shows the fidelity for obtaining the Dicke state, $P_{\rm mid}^{\rm QA} = \vert \langle \psi(T/2) \vert S^z=0 \rangle|^2$ and the generalized squeezing,  $\xi^2_{\rm QA} = \frac{\Delta S_z^2}{N/4}$ at $t=T/2$ for a quantum annealing protocol that starts at $t=0$ and stops at $t=T/2$; here $T$ is set to $2$. CD driving leads to an enhancement of both $P_{\rm mid}^{\rm QA}$ and $\xi^2_{\rm QA}$ and it works best in the low-$J$ regime. The efficacy of CD driving decreases with increasing $J$ for this protocol. On the other hand, in the absence of CD driving, greater fidelity and better spin squeezing is achieved with increasing $J$. (c)-(d) shows the maximum fidelity for obtaining the Dicke state, $P_{\rm mid}^{\rm max} = \vert \langle \psi(t=(n+1/2)T) \vert S^z=0 \rangle|^2$ and the optimal (minimum) value of the generalized squeezing,  $\xi^2_{\rm opt} = \frac{\Delta S_z^2 (t=(n+1/2)T) }{N/4}$ over the first 40 drive cycles. CD driving leads to an enhancement of both $P_{\rm mid}^{\rm max}$ and $\xi^2_{\rm opt}$ compared to $P_{\rm mid}^{\rm QA}$ and $\xi^2_{\rm QA}$ respectively. However, the performance of these protocols exhibits a non-monotonic behavior with increasing $J$. Intriguingly,  the CD1 protocol performs better (worse) than the CD2 protocol when $J \lesssim 1$ ($J \gtrsim 1$). These results have been obtained for $N=100$. In the absence of CD driving, $P_{\rm mid}^{\rm max}$ shows a weak dependence on $J$, but  $\xi^2_{\rm opt}$ monotonically decreases with $J$ for $J \gtrsim 0.5$.}
    \label{fig:suppfig2}
\end{figure}

\subsection{Dicke State Preparation Protocol}
\label{subsec:Dicke}
In this sub-section, we discuss the efficacy of CD driving for preparing Dicke states. In this work, we analyze the Hamiltonian :
\begin{equation}
    H = -\bigl(1-\lambda(t)\bigr) \sum_{i=1}^N \sigma_i^x + \lambda(t) \frac{J}{N} \sum_{i,j=1}^N \sigma_i^z \sigma_j^z,
\label{eq:modelsupp}
\end{equation}
where $\lambda(t)=\sin^2{\left[\left(\frac{\pi}{2}\right)\sin^2{}\left(\frac{\pi t}{T}\right)\right]}$. An adiabatic state preparation protocol would lead to the generation of spin-squeezed Dicke states at $t=T/2$. Firstly, following the usual quantum annealing approach~\cite{claeys2019floquet}, we analyze the effect of CD driving on the state preparation fidelity by computing both $P_{\rm mid}^{\rm QA} = \vert \langle \psi(T/2) \vert S^z=0 \rangle|^2$ and $\xi^2_{\rm QA} = \frac{\Delta S_z^2}{N/4}$ at $t=T/2$. We find that the CD driving is extremely effective at low values of $J$, where a very high value of $P_{\rm mid}^{\rm QA}$ is obtained; this value decreases with larger $J$ for the CD driving. Without CD driving on the other hand, $P_{\rm mid}^{\rm QA}$ is very small at low values of $J$ and it increases with increasing $J$. An analogous behavior is seen for the spin-squeezing parameter, $\xi^2$. In all of these cases, CD driving always leads to an enhancement of $P_{\rm mid}^{\rm QA}$ and the generation of greater spin-squeezing. Our results are shown in Fig.~\ref{fig:suppfig2} (a)-(b).\\

We have already noted in the main text that the standard quantum annealing procedure may not be the most optimal route to prepare Dicke states. Instead, a better fidelity might be obtained by running the Floquet protocol for a few cycles. In order to investigate this, we computed the maximum value of $P_{\rm mid} = \vert \langle \psi((n+1/2)T) \vert S^z=0 \rangle|^2$ (dubbed $P_{\rm mid}^{\rm max}$) and the optimal (minimum) value of $\xi^2 = \frac{\Delta S_z^2\big((n+1/2)T\big)}{N/4}$ (dubbed $\xi^2_{\rm opt}$) over the first 40 drive cycles. Firstly, CD driving always leads to a greater value of $P_{\rm mid}^{\rm max}$ and a lower value of $\xi^2_{\rm opt}$ compared to $P_{\rm mid}^{\rm QA}$ and $\xi^2_{\rm QA}$ respectively. Interestingly, however, the CD driving protocols exhibit a non-monotonic dependence on $J$. Furthermore, the CD1 protocol is more (less) effective than the CD2 protocol for state preparation when $J \lesssim 1$ ($J \gtrsim 1$). Our results are shown in Fig.~\ref{fig:suppfig2} (c)-(d), and they clearly establish that employing a Floquet CD protocol leads to a better fidelity for state preparation.

\section{Eigenstate Localization}
\label{sec:Localization}

In this section, we provide a brief overview of the measures used to characterize the localization of the Floquet eigenstates $|\mu_n\rangle$. However, before we describe the localization measures, let us first define the generalized SU$(2)$ spin-coherent states, as follows:
\begin{equation}
\vert \theta, \phi \rangle = \exp\left[i \theta(S^x \sin(\phi) - S^y \cos(\phi))\right] \vert S^z = N/2 \rangle.
\end{equation}
These states admit a natural representation as points on the surface of a Bloch sphere parameterized by the variables $\theta$, and $\phi$.\\

We now introduce the Husimi Q-distribution $Q_n(\theta,\phi)$, which is a quasi-probability distribution that helps us represent the phase space distribution of Floquet eigenstates, $\vert \mu_n \rangle$. 
\begin{equation}
\ Q_n (\theta, \phi)=\vert \langle \theta, \phi \vert \mu_n \rangle \vert^2.
\end{equation}

This distribution measures the overlap of the Floquet eigenstate,  $\vert \mu_n \rangle$ with the spin coherent states defined above. Being a quasi-probability distribution that is a measure of the overlap between the states, we can use $Q_n(\theta,\phi)$ to define the Wehrl entropy and inverse participation ratio, over the phase space~\cite{wang2023statistics}. Using the definition of the Von Neumann entropy on the Husimi distribution, we can define the Wehrl entropy as follows:
\begin{equation}
\ S = -\sum_i p_i \ln{p_i} \sim -\frac{N+1}{4\pi}\int_{0}^{2\pi}\int_{0}^{\pi} Q_n(\theta,\phi) \ln{Q_n(\theta,\phi)} \sin{\theta}d\theta d\phi = S_w
\end{equation}

We now use the Wehrl entropy, $S_w$ to define the eigenstate localization by defining the Wehrl entropy Localization measure $\mathcal{L}$: 
\begin{equation}
\mathcal{L} = \exp(S_w)/(N+1)
\end{equation}
The eigenstate averaged localization measure takes the value of $0.655$ in a completely ergodic system for large $N$~\cite{yan2024addendum}.\\

Furthermore, we examine the Inverse Participation Ratio (IPR), which is another localization measure. The IPR for a Floquet eigenstate $\vert \mu_n\rangle$, $I_n$: 
\begin{equation}
\ I_n =\frac{(N+1)^2}{4\pi}\left[\int_{0}^{2\pi}\int_{0}^{\pi} Q_n^2(\theta,\phi)  \sin{\theta}d\theta d\phi\right]^{-1}.
\end{equation}
The IPR $I_n$ is very small $(\sim 0)$ for extremely localized eigenstates and takes a value of $1$ for completely delocalized eigenstates~\cite{wang2023statistics}. Without CD driving, the eigenstate-averaged IPR takes a large value ($\sim 0.5$) in the ergodic regime; this is similar to the results obtained for the kicked top~\cite{wang2023statistics}. The value of the IPR is considerably reduced by CD driving. Our results are shown in Fig.~4 of the main text.

\section{Spin Squeezing}
\label{sec:Squeeze}

Spin squeezing offers a powerful and experimentally feasible method to detect multipartite entanglement. In this work, we quantitatively characterize spin squeezing using the parameter, $\xi^2$~\cite{opatrny2016counterdiabatic}:
\begin{equation}
\xi^2 = \frac{\Delta \hat{S}_z^2}{N/4}
\end{equation}
where $\xi^2 < 1$ indicates spin-squeezing and multipartite entanglement. In this section, we examine spin-squeezing and its relation to entanglement in greater detail.\\

Our model is described by collective spin operators $\hat{S}_x$, $\hat{S}_y$, and $\hat{S}_z$, where each operator represents the sum of individual spin components:
\[
\hat{S}_\alpha = \sum_{i=1}^N \hat{s}_\alpha^{(i)} \quad \text{for } \alpha = x, y, z.
\]
The total spin operator is $\hat{\mathbf{S}}^2 = \hat{S}_x^2 + \hat{S}_y^2 + \hat{S}_z^2$, and for symmetric states of $N$ spin-$\tfrac{1}{2}$ particles, the total spin quantum number is $S = N/2$. Therefore, such states satisfy:
\[
\langle \hat{\mathbf{S}}^2 \rangle = S(S + 1) = \frac{N}{2} \left( \frac{N}{2} + 1 \right).
\]

A fundamental property of separable (i.e., unentangled) states is that they obey the inequality:
\[
\langle \hat{S}_x^2 \rangle + \langle \hat{S}_y^2 \rangle \leq \frac{N}{2} \left( \frac{N}{2} + \frac{1}{2} \right),
\]
which, when combined with the identity $\langle \hat{\mathbf{S}}^2 \rangle = \langle \hat{S}_x^2 \rangle + \langle \hat{S}_y^2 \rangle + \langle \hat{S}_z^2 \rangle$, leads to the bound:
\[
\mathrm{Var}(\hat{S}_z) = \langle \hat{S}_z^2 \rangle - \langle \hat{S}_z \rangle^2 \geq \frac{N}{4}.
\]
For target states with $\langle \hat{S}_z \rangle = 0$, such as spin-squeezed and Dicke states, this implies:
\[
\mathrm{Var}(\hat{S}_z) = \langle \hat{S}_z^2 \rangle \geq \frac{N}{4}.
\]

Spin squeezing refers to the reduction of quantum fluctuations of one spin component—such as $\hat{S}_z$—below this standard quantum limit:,
\[
\xi^2 = \frac{\Delta \hat{S}_z^2}{N/4} < 1.
\]
This reduction is accompanied by a corresponding increase in fluctuations in the orthogonal spin component, in accordance with the uncertainty relation. Such squeezing cannot occur in separable states, and it is a direct signature of multipartite entanglement~\cite{sorensen2001entanglement,sorensen2001many}. \\

In the extreme case of the symmetric Dicke state $|S_z = 0\rangle$, the spin projection, $\mathrm{Var}(\hat{S}_z) = 0$, and the transverse fluctuations saturate the total spin:
\[
\langle \hat{S}_x^2 \rangle + \langle \hat{S}_y^2 \rangle = \frac{N}{2} \left( \frac{N}{2} + 1 \right).
\]
This maximally violates the separability bound and indicates the presence of strong multipartite entanglement~\cite{santra2024squeezing}.\\

\section{Level Spacing Statistics}
\label{sec:LevelStat}

Spectral statistics is a very useful and powerful tool, that can be used to diagnose the nature of quantum many-body systems. In particular, the level spacing ratio $r_n$ is a simple measure, that is often used to study these eigenstates, as it provides a quantitative measure of chaos~\cite{pal2010many,atas2013distribution,d2014long}. We define $r_n$ as follows,

\begin{equation}
r_n = \frac{{\rm min}(d_n,d_{n+1})}{{\rm max}(d_n,d_{n+1})},
\end{equation}

where, $d_n = \epsilon^F_{n+1}-\epsilon^F_{n}$ is the Floquet level-spacing and $\epsilon^F_{n}$ represent the quasienergies of the Floquet Hamiltonian, $H_F$. As noted in the main text, the quasienergies and the Floquet Hamiltonian are obtained from the Floquet operator, $U_F = \mathcal{T} \exp[-i \int dt H(t)] = \exp(- i H_F T) = \sum_n \exp(-i \epsilon^F_{n} T) \vert \phi_n \rangle \langle \phi_n \vert $.\\

For integrable systems characterized by Poisson statistics, the probability distribution of the level spacing ratios $r_n$, $P(r)$ takes the form

\begin{equation}
P(r) = \frac{2}{(1+r)^2},
\end{equation}

and $\langle r\rangle \sim 0.386$~\cite{pal2010many,atas2013distribution,d2014long}. For chaotic systems following the Wigner-Dyson distribution on the other hand, we obtain 

\begin{equation}
P(r) = \frac{27}{4}\frac{r+r^2}{(1+r+r^2)^{5/2}},
\end{equation}
such that $\langle r\rangle \sim 0.531$~\cite{pal2010many,atas2013distribution,d2014long}.

\bibliographystyle{apsrev4-1}
\bibliography{ref}